\RequirePackage{fix-cm}
\documentclass{svjour3}             
\smartqed  
\usepackage{mathptmx}      
\usepackage{graphicx}
\usepackage{wasysym}
\usepackage{natbib}
\usepackage[urlcolor=blue]{hyperref}
\hypersetup{colorlinks, linkcolor={black}, citecolor={blue}}
\journalname{Celestial Mechanics and Dynamical Astronomy}

\begin{document}

\title{Mercury's resonant rotation from secular orbital elements}


\author{Alexander Stark \and J\"urgen Oberst \and  Hauke Hussmann 
}
\institute{A. Stark \and J. Oberst \and H. Hussmann \at
              DLR, Institute of Planetary Research\\
							Rutherfordstr. 2, D-12489 Berlin\\
              Tel.: +49-30-67055696, Fax: +49-30-67055402 \\
							\email{Alexander.Stark@dlr.de} \\\\
							J. Oberst\\
							Moscow State University for Geodesy and Cartography\\
							RU-105064 Moscow, Russia
}

\date{The final publication is available at \href{http://dx.doi.org/10.1007/s10569-015-9633-4}{http://link.springer.com}}
%

\maketitle

\begin{abstract}
We used recently produced Solar System ephemerides, which incorporate two years of ranging observations to the MESSENGER spacecraft, to extract the secular orbital elements for Mercury and associated uncertainties. 
As Mercury is in a stable 3:2 spin-orbit resonance these values constitute an important reference for the planet's  measured rotational parameters, which in turn strongly bear on physical interpretation of Mercury's interior structure.
In particular, we derive a mean orbital period of $(87.96934962 \pm 0.00000037)\,\textrm{days}$ and (assuming a perfect resonance) a spin rate of $(6.138506839\pm 0.000000028){}^{\circ}/\textrm{day}$. 
The difference between this rotation rate and the currently adopted rotation rate \citep{archinal11} corresponds to a longitudinal displacement of approx. $67\,\textrm{m}$ per year at the equator.
Moreover, we present a basic approach for the calculation of the orientation of the instantaneous Laplace and Cassini planes of Mercury. The analysis allows us to assess the uncertainties in physical parameters of the planet, when derived from observations of Mercury's rotation.

\end{abstract}

\section{Introduction}
\label{intro}
Mercury's orbit is not inertially stable but exposed to various perturbations which over long time scales lead to a chaotic motion \citep{laskar89}. The short-term (about few thousand years) evolution of the orbit can be approximated by a secular contribution to the orbital elements. Most prominent is the precession of the pericenter of Mercury's orbit, which was also an important test of Einstein's theory of general relativity \citep{einstein15}. Due to the Sun's torque on the asymmetric mass distribution of Mercury, the rotation of Mercury is strongly coupled to its evolving orbit. Radar observations \citep{pettengill65} revealed that Mercury's rotation period is about 59 days and in a stable 3:2 resonance with its orbital period \citep{peale65,colombo65}. 
More recently \cite{margot07} have used an Earth-based radar-speckle  correlation technique to precisely measure the physical libration amplitude and the obliquity of Mercury. 
By interpretation of these measurements in terms of physical parameters of the planet - following the idea of the Peale experiment \citep{peale76,peale81} - the authors concluded that Mercury's core is at least partially molten \citep{margot07,margot12}.
\begin{figure}
\centering
 \includegraphics[width=1\textwidth]{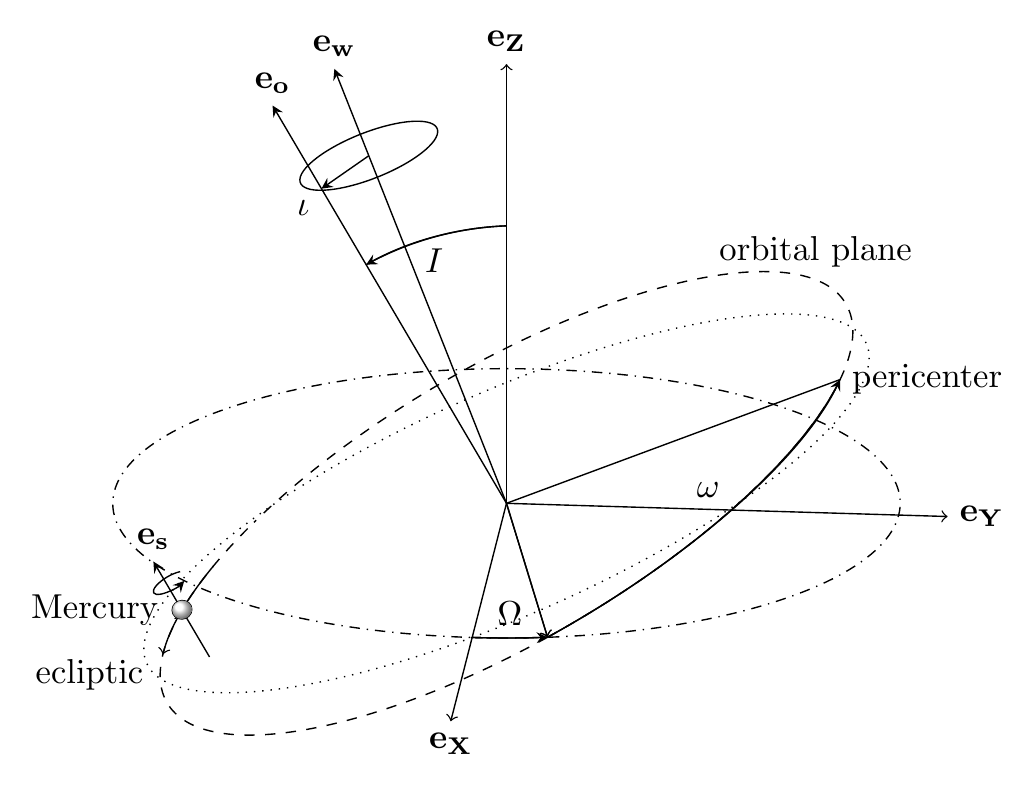}
  \caption{The unit vectors $\mathbf{e_{X,Y,Z}}$ denote the International Celestial Reference Frame (ICRF). Mercury's orbital plane is illustrated by a dashed ellipse and its orientation by the vector $\mathbf{e_o}$. The ecliptic and the $\mathbf{e_X}$-$\mathbf{e_Y}$ plane of the ICRF are given by dotted and dash-dotted ellipses, respectively. The Laplace plane normal is indicated by $\mathbf{e_w}$ and Mercury's spin axis by $\mathbf{e_s}$. The figure is based on numbers given in Tab. 
	\ref{tab:elem} at the J2000.0 epoch.}
	\label{fig:ref}
\end{figure}

With the MESSENGER space probe (MErcury Surface, Space ENvironment, GEochemistry, and Ranging) having entered orbit around Mercury in March 2011, the observational data of Mercury have greatly improved. Further, new Solar System ephemerides which incorporate two years of ranging and Doppler tracking observations to MESSENGER were produced. For the interpretation of the observations of Mercury's rotation performed by instruments on MESSENGER, precise knowledge of the resonant rotation parameters of Mercury is mandatory. In fact, the resonant spin rate, currently adopted in the rotation model of Mercury, dates back to the first IAU report \citep{davies80}.

In this work we provide updated reference values for Mercury's rotation assuming the perfectly resonant rotation model based on the most recent planetary ephemerides. These values serve as a basis for the interpretation of the rotational parameters of Mercury, which are proposed to be measured with high precision \citep{stark15a}.

\section{Secular orbital elements of Mercury}
Recently, new Solar System ephemerides DE432 from the Jet Propulsion Laboratory (W. M. Folkner, personal communication, 2014) and INPOP13c from the Institut de M\'ecanique C\'eleste et de Calcul des \'Eph\'em\'erides \citep{fienga14} were produced. Besides other improvements these ephemerides incorporate updates to the orbit of Mercury. Both ephemerides although different in their production process and covered time span led to identical results in our calculations. We concentrate here on the DE432 ephemeris and give the orbital elements derived from the INPOP13c ephemeris in appendix \ref{app:ref}.

The DE432 ephemeris covers a time span of approximately 1,000 years (1 January 1550 to 1 January 2550).
In this time span we derived the osculating Keplerian orbital elements of Mercury from state vectors given with respect to the Sun-centered International Celestial Reference Frame (ICRF). We used a time step of 7 days and set the gravitational parameter of the Sun to $GM_{\astrosun}=132,712,440,041.9394\,\textrm{km}^3/\textrm{s}^2$ \citep{folkner14}. For the calculation of the osculating orbital elements standard techniques were used \citep{bate70}.
In order to obtain the secular parts of the elements we decomposed the osculating orbital elements in a quadratic polynomial and a sum of periodic terms
\begin{equation}
  x(t) = x_0 + x_1 t/\textrm{cy} + x_2 (t/\textrm{cy})^2 + \sum_i A_i\cos(\nu_i\,t +\phi_i)\,,
	\label{eq:x}
\end{equation}
where $x$ stands for a Keplerian orbital element $a$, $e$, $I$, $\Omega$, $\omega$, $M$, being semi-major axis, eccentricity, inclination, longitude of ascending node, argument of pericenter, and mean anomaly, respectively. The time $t$ is measured in Julian centuries (cy) from the J2000.0 epoch. Higher order terms in the polynomial were discarded because their estimated uncertainty exceeded the actual value by many orders of magnitude. The periodic terms are characterized by their amplitude $A_i$, frequency $\nu_i$, and phase $\phi_i$. We list these values for ten highest amplitudes of each orbital element in appendix \ref{app:freq}.

The decomposition of the osculating orbital elements time series into the form of Eq. \ref{eq:x} was performed with the help of the frequency mapping tool FAMOUS \footnote{F. Mignard, OCA/CNRS, ftp://ftp.obs-nice.fr/pub/mignard/Famous}. This is done because a simple least-squares fit may lead to biased results given the fact that the variations of the orbital elements are in first order periodic and not random. At least 50 frequencies were identified and subtracted from the variation of the orbital elements. The variance of the periodic variations $\sigma^2_x$ was used to estimate an uncertainty for the coefficients. Thus, orbital elements with relatively high periodic variations receive higher error bars. The uncertainties of the secular coefficients $x_1$ and $x_2$ were derived by considering the maximal slope and curvature of the polynomial within the interval $[-\sigma_x,\sigma_x]$ and a time span of 1000 years. The resulting values are given in Table \ref{tab:elem}.
\begin{table}
\begin{tabular}{c r r r }
\hline\noalign{\smallskip}
         & $x_0$ & $x_1$ & $x_2$ \\
				\noalign{\smallskip}\hline\noalign{\smallskip}
	  $a/(10^6\,\textrm{km})$ & $57.90909$ &     $0.002\times 10^{-6}$ & $    -0.002\times 10^{-6}$ \\
		                    &  $\pm 0.00011$ & $\pm 22.34\times 10^{-6}$ & $\pm 4.45\times 10^{-6}$ \\
		$e$  &      $0.2056317$ &    $20.4\times 10^{-6}$ &      $-20\times 10^{-6}$ \\
		     &  $\pm 0.0000071$ & $\pm 1.4\times 10^{-6}$ & $\pm 290\times 10^{-6}$ \\
		$I/{}^{\circ}$  & $28.552197$ &     $0.0048464$ &    $-9.8\times 10^{-6}$ \\
		     &         $\pm 0.000036$ & $\pm 0.0000073$ & $\pm 1.5\times 10^{-6}$ \\
		$\Omega/{}^{\circ}$ & $10.987971$ &     $-0.032808$ &   $-12.3\times 10^{-6}$ \\
		     &             $\pm 0.000099$ & $\pm 0.000020$ & $\pm 4.0\times 10^{-6}$ \\
		$\omega/{}^{\circ}$ & $67.5642$ &    $0.18861$ &   $  -3\times 10^{-6}$ \\
		     &             $\pm 0.0020$ & $\pm 0.00040$ & $\pm 80\times 10^{-6}$  \\
		$M/{}^{\circ}$  & $174.7948$ &     $149472.51579$ &   $   8\times 10^{-6}$ \\
		     &         $\pm 0.0032$ & $\pm 0.00063$ & $\pm 126\times 10^{-6}$ \\
\end{tabular}
	\caption{Secular Keplerian orbital elements of Mercury as derived from the DE432 ephemeris at epoch J2000.0, given with respect to ICRF (see Fig. \ref{fig:ref}).}
	\label{tab:elem}
\end{table}

In order to demonstrate the convergence of the method we increased the number of frequencies to 100 and found only changes below 2\% of the uncertainties of the polynomial coefficients. 
For further verification of our approach we calculated orbital elements with respect to the ecliptic at J2000.0 (see appendix \ref{app:ref}) and compared our results with those published by \cite{standish13}. Beside the secular parts of the inclination and longitude of ascending node our values and their uncertainties are consistent with the published values. The discrepancy we found in $I_1$ and $\Omega_1$ may result from the fact that we consider the quadratic term, which is significant for these elements. Comparison with other literature values \citep{margot09,noyelles12,noyelles13} shows excellent agreement with our values for these orbital elements.

Additionally, we calculated the precession of the pericenter of Mercury. Note that the secular rates are strongly dependent on the selected reference frame. We used the mean orbital plane of Mercury at J2000.0 (see Sec. \ref{sec:mop}) as reference frame and found a precession of $575.3\pm1.5\,\textrm{arc sec}/\textrm{cy}$ (see appendix \ref{app:ref}). Again this is in a very good agreement to the literature value of $(5600.73 - 5025)\,\textrm{arc sec}/\textrm{cy}=575.73\pm0.41\,\textrm{arc sec}/\textrm{cy}$ \citep[computed from][p.199]{weinberg72}.

Another method to obtain the secular orbital elements involve the usage of a secular potential and integration of the averaged differential equation of Mercury's motion. However, such a method neglects the mutual interaction of the perturbing planets and is not appropriate for precise interpretation of spacecraft data \citep{yseboodt06}. More details on averaging methods can be found in e.g., \cite{sanders07}.

\subsection{Mean orbital period}
The mean period of the orbit is defined as $T_{\textrm{\small orbit}}=2\pi/n_0$, where $n_0$ is the mean motion of Mercury. We can derive $n_0$ from the first order term of the mean anomaly $M=M_0 + M_1 t = n_0(t_0+t)$. The time $t_0$ is the elapsed time at J2000.0 since the last pericenter passage. Using the values in Tab. \ref{tab:elem} we derive
\begin{eqnarray}
	n_0  &=& M_1 = (4.092334450 \pm 0.000000017){}^{\circ}/{\textrm{day}} 
	\label{eq:n0}\\
	t_0  &=& M_0/M_1 = (42.71274 \pm 0.00077)\,\textrm{day}	\label{eq:t0}\\
	T_{\textrm{\small orbit}} &=& 360^{\circ}/M_1 = (87.96934962 \pm 0.00000037)\,\textrm{day}\,.	\label{eq:Torbit}
\end{eqnarray}
 In order to check the derived value of $n_0$ we used Kepler's third law $n_0=\sqrt{GM_{\astrosun}/a_0^3}$ and found $4.092343\pm 0.000083\,{}^\circ\!/\textrm{day}$.  This value is consistent with Eq. \ref{eq:n0} but has an error larger by two orders of magnitude.

\subsection{Mean orbital plane}
\label{sec:mop}
From the secular parts of the orbital elements $\Omega$ and $I$ we can calculate the mean normal vector of the orbital plane by
\begin{eqnarray}
	\label{eq:n}
  \mathbf{e_o} &=& \sin\Omega\,\sin I \, \mathbf{e_X} - \cos\Omega\,\sin I \,\mathbf{e_Y} + \cos I \,\mathbf{e_Z}\\
		\label{eq:orbit}
	 &=& \cos\alpha^{\textrm{orbit}}\,\cos \delta^{\textrm{orbit}} \,\mathbf{e_X} + \sin\alpha^{\textrm{orbit}}\,\cos \delta^{\textrm{orbit}} \,\mathbf{e_Y} + \sin \delta^{\textrm{orbit}}\,\mathbf{e_Z}\,,
\end{eqnarray}
where $\mathbf{e_{X,Y,Z}}$ denote the orientation of the ICRF (see also Fig. \ref{fig:ref}). Comparing Eq. 5 and 6 we find the right ascension and declination of the orbit pole to be $\alpha^{\textrm{orbit}}=\Omega - \pi/2$ and $\delta^{\textrm{orbit}}=\pi/2-I$. At J2000.0 the values are $\alpha^{\textrm{orbit}}_{0} = (280.987971\pm0.000099){}^{\circ}$ and $\delta^{\textrm{orbit}}_{0}=(61.447803\pm 0.000036){}^{\circ}$, respectively. From the secular components of $\Omega$ and $I$ we can directly derive the first order precession rates of the orbit pole $\alpha_1^{\textrm{orbit}} = \Omega_1 = (-0.032808\pm 0.000020){}^{\circ}/\textrm{cy}$ and $\delta_1^{\textrm{orbit}} = -I_1 = (-0.0048464\pm 0.0000073){}^{\circ}/\textrm{cy}$. 
It should be noted that the precession of the orbit normal is treated here as a secular variation in inclination and longitude of ascending node, which is justified by the long period of the precession. By that reason the given description of the mean orbital plane is strictly valid only for the time span of the ephemeris, i.e. about $\pm 500$ years around the J2000.0 epoch.
The error bars on the orbit pole orientation and precession rates were obtained through propagation of the uncertainties in the orbital elements. Note that the derived values are in agreement with the findings of \cite{margot09} with $\alpha^{\textrm{\small orbit}}=280.9880{}^{\circ}-0.0328{}^{\circ}t/{\textrm{cy}}$ and $\delta^{\textrm{\small orbit}} =61.4478{}^{\circ}-0.0049{}^{\circ}t/{\textrm{cy}}$, where DE408 ephemeris and a period of 200 year was used. 

\subsection{Laplace plane}
\begin{figure}
 \centering
 \includegraphics[width=0.7\columnwidth]{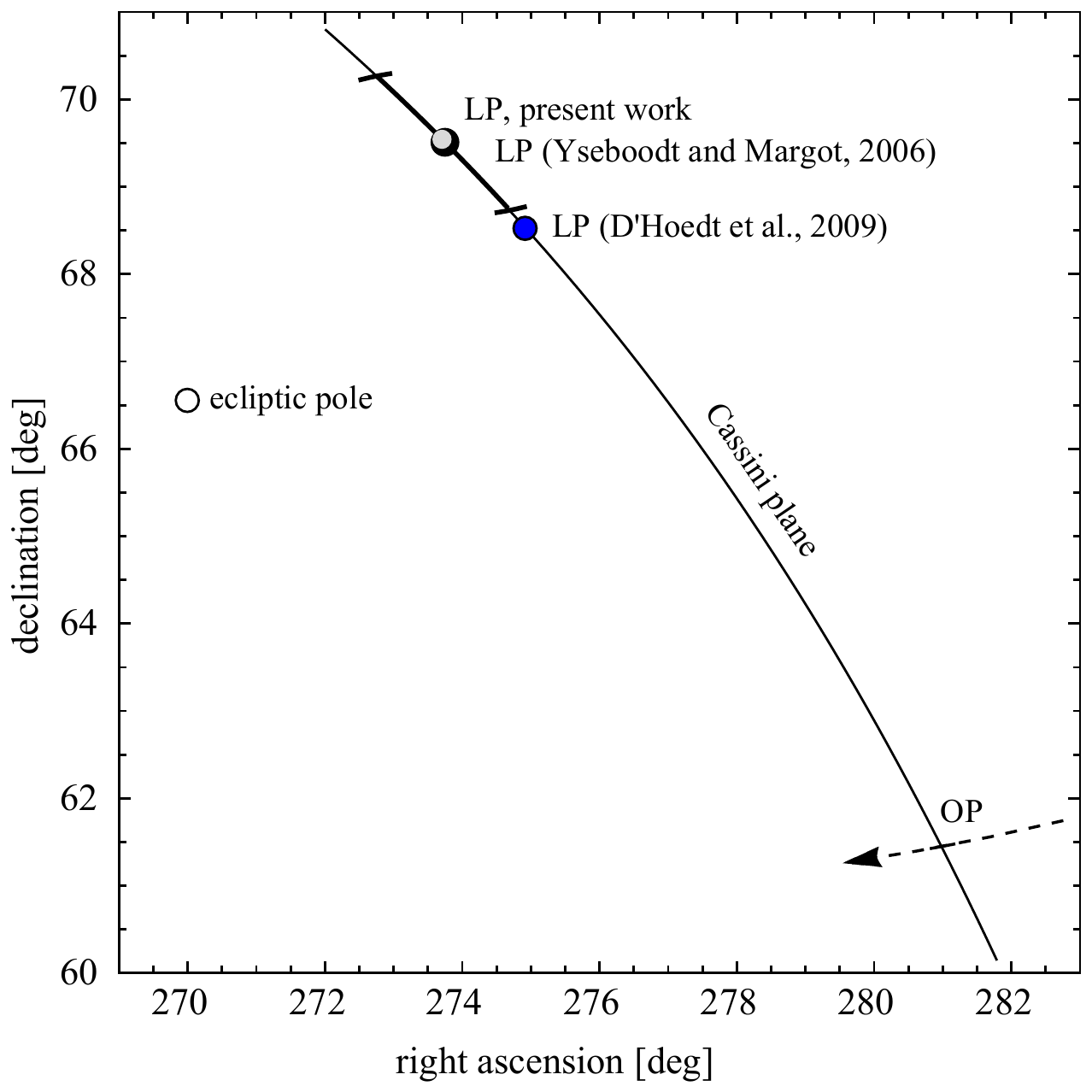}
  \caption{Orientation of the orbit (OP) and the Laplace plane (LP, black disk with error bars) normal at J2000.0 epoch with respect to the ICRF. 
The precession of the orbit pole around the instantaneous Laplace pole is indicated by a dashed arc. The Laplace plane orientation of \cite{yseboodt06} (grey
disk) overlaps with the values derived in this work. Note that the longitude of the Laplace pole is given incorrect in \citep{dhoedt09} and was corrected in \citep{noyelles12}. The figure shows the corrected position (blue disk).}
	\label{fig:loc}
\end{figure}

The other planets of the Solar System exert a torque on the orbital plane of Mercury, which leads to a quasi-periodic precession of the orbit normal. Further, the plane to which the inclination of Mercury remains constant, i.e., the Laplace plane, also undergoes slow variations \citep{noyelles12}. Several attempts have been made to calculate the orientation of the Laplace plane normal \citep{yseboodt06,peale06,dhoedt09}, each of them leading to different results in the Laplace pole position and the precession period (see Fig. \ref{fig:loc}).

The concept of the "instantaneous" Laplace plane was proposed for Mercury by \cite{yseboodt06} to derive an approximate Laplace plane valid for several thousand years. Note that without additional assumptions the instantaneous precession vector $\mathbf{w}$ is only constrained to a line. In order to overcome the ambiguity in the instantaneous Laplace plane either a fit to the ephemeris \citep{yseboodt06} or some additional assumptions \citep{peale06,dhoedt09} are used. 

Here we introduce a concept of an instantaneous Laplace plane, which removes the ambiguity in its instantaneous orientation and  precession frequency. As the only assumption we require the instantaneous Laplace plane to be invariable, i.e. $\dot{\mathbf{w}}\equiv 0$. Note that a similar concept was suggested by \cite{yseboodt11}. When an instantaneous Laplace plane is considered it is also important to clarify in which time period it should be instantaneous. The free precession period of Mercury is in the order of 1000 years \citep{peale05}. This means that the rotation axis will not be affected by the short period (on the order of decades) changes in the orientation of the orbit normal, but will follow the changes at long periods due to adiabatic invariance \citep{peale05}. In order to obtain an instantaneous Laplace plane, which is relevant for Mercury's spin, we can neglect all periodic variations and consider only the secular terms.

The general equation for the precession around an axis $\mathbf{w}$ is 
\begin{equation}
\mathbf{w} \times \mathbf{e_o} = \mathbf{\dot{e}_o}\,.
\label{eq:wn}
\end{equation}
The precession vector $\mathbf{w}$ is given by $\mathbf{w}=-\mu\,\mathbf{e_w}$, where $\mathbf{e_w}$ is the orientation of the Laplace plane and $\mu$ the precession rate.
First we multiply both sides of Eq. \ref{eq:wn} with $\mathbf{e_o}$ and obtain
\begin{equation}
  \mathbf{w} = (\mathbf{e_o}\times\mathbf{\dot{e}_o}) - \mu \cos \iota\, \mathbf{e_o}\,,
	\label{eq:w}
\end{equation}
where we used $\mathbf{e_o}\cdot\mathbf{w}=-\mu\cos\iota$ and
$\iota$ is the inclination of Mercury's orbit with respect to the Laplace plane. 
In order to constrain the instantaneous orientation of the Laplace plane we have to find an instantaneous value for $\mu \cos \iota$.
By differentiating Eq. \ref{eq:wn} and the requirement $\dot{\mathbf{w}}\equiv 0$ we obtain
\begin{equation}
  \mathbf{\ddot{e}_o} + \mu^2\,\mathbf{e_o} = -\mu \cos \iota\,\mathbf{w}\,,
	\label{eq:diffn}
\end{equation}
where we make use of $\mathbf{w}\times\mathbf{\dot{e}_o}=\mathbf{w}\times(\mathbf{w} \times\mathbf{e_o})=-\mu\cos\iota\mathbf{w}-\mu^2\mathbf{e_o}$.
The differential equation Eq. \ref{eq:diffn} describes a regular (uniform) rotation of $\mathbf{e_o}$ around $\mathbf{e_w}$ with the frequency $\mu$. Following the concept of the instantaneous Laplace plane Eq. \ref{eq:diffn} is only fulfilled with a unique $\mathbf{w}$ for a specific time $t$. By multiplying Eq. \ref{eq:diffn} with $\mathbf{e_o}$ and using Eq. \ref{eq:wn} we can find
\begin{equation}
  \mu \cos \iota = \frac{\mathbf{\dot{e}_o}\cdot(\mathbf{e_o}\times\mathbf{\ddot{e}_o})}{|\mathbf{\dot{e}_o}|^2}\,.
	\label{eq:mucosi}
\end{equation}
By using Eq. \ref{eq:n} for $\mathbf{e_o}$ and $\mathbf{\ddot{e}_o}$ we can obtain the values of $\mu \cos \iota$ as it would be in a regular form and by that an expression for the instantaneous Laplace plane. The combination of Eq. \ref{eq:w} and \ref{eq:mucosi} gives the instantaneous Laplace plane orientation
\begin{equation}
  \mathbf{w} = (\mathbf{e_o}\times\mathbf{\dot{e}_o}) - \frac{\mathbf{\dot{e}_o}\cdot(\mathbf{e_o}\times\mathbf{\ddot{e}_o})}{|\mathbf{\dot{e}_o}|^2}\mathbf{e_o}\,.
	\label{eq:wfin}
\end{equation}
Following the formalism of \citet{peale06} Eq. \ref{eq:wfin} can be expressed as
\begin{eqnarray}
  \mathbf{w} = & \left(\dot{I}\cos \Omega+\left(w_z - \dot{\Omega}\right)\tan I \sin \Omega\right)\mathbf{e_X} +\nonumber\\ & + 
	             \left(\dot{I}\sin \Omega-\left(w_z - \dot{\Omega}\right)\tan I \cos \Omega\right)\mathbf{e_Y} + 
							 w_z \mathbf{e_Z}
  \label{eq:wfull}
\end{eqnarray}
and $w_z$ given by 
\begin{equation}
  w_z = \dot{\Omega}+\frac{(\ddot{I}\dot{\Omega}-\dot{I}\ddot{\Omega})\sin I+ \dot{\Omega} \dot{I}^2 \cos I}{\dot{I}^2+(\dot{\Omega}\sin I)^2}\cos I\,.
\end{equation}
The instantaneous Laplace pole given by Eq. \ref{eq:wfin} is practically equivalent to the fit of the ephemeris to a cone, performed by \cite{yseboodt06}.

Using Eq. \ref{eq:wfin} we calculate the coordinates of the Laplace pole at J2000.0 to $\alpha^{\textrm{\small LP}}_{0} = (273.8\pm 1.0){}^{\circ}$ and $\delta^{\textrm{\small LP}}_{0} = (69.50\pm 0.77){}^{\circ}$ in the ICRF. It should be noted that the covariance $\textrm{Cov}(\alpha^{\textrm{\small LP}}_{0},\delta^{\textrm{\small LP}}_{0})=-(0.77{}^{\circ})^2$ is very high, indicating a high correlation $(-99.8\%)$ of the right ascension and declination values. The instantaneous precession rate is $\mu=(0.00192\pm 0.00018)/\textrm{cy}$, $T_{\textrm{\small LP}} = (327300\pm32000)\,\textrm{years}$ and $\iota = (8.58\pm0.84){}^{\circ}$. 

However, if Mercury is in a Cassini state (see Sec. \ref{sec:cas}) the determination of the polar moment of inertia from the obliquity is not largely affected by the uncertainties of the Laplace pole \citep{yseboodt06,peale06}. Using our formalism we can derive the relevant quantities and their errors
\begin{eqnarray}
  \mu \sin \iota =& (2.8645\pm0.0016)\times 10^{-6}/\textrm{years}\\
	\mu \cos \iota =& (18.98\pm 1.83)\times 10^{-6}/\textrm{years}\,.
\end{eqnarray}
The correlation between $\mu \sin \iota$ and $\mu \cos \iota$ is very low $\textrm{Corr}(\mu \sin \iota,\mu \cos \iota)=-10^{-3}$.

\section{Mercury's rotation}
\subsection{Rotation model}
The rotation model of a celestial body consists of a set of values defining its orientation as a function of time with respect to a reference frame. Here we recall briefly the IAU convention of a rotation model for Mercury \citep[e.g.][]{archinal11}. 

The orientation of Mercury's spin axis is described by the right ascension $\alpha$ and declination $\delta$ coordinates of the intercept of the spin axis vector $\mathbf{e_s}$ with the celestial sphere. The spin axis vector $\mathbf{e_s}$ with respect to the ICRF is given by
\begin{equation}
  \mathbf{e_s} = \cos\alpha\,\cos \delta \,\mathbf{e_X} +\sin\alpha\,\cos \delta \,\mathbf{e_Y} + \sin \delta\,\mathbf{e_Z}\,,
	\label{eq:s}
\end{equation}
where $\mathbf{e_{X,Y,Z}}$ denotes the ICRF.
The orientation at J2000.0 epoch is denoted by $\alpha_0$ and $\delta_0$ and the first order precession rates are $\alpha_1$ and $\delta_1$. The rotational axis is consequently given by
\begin{eqnarray}
     \alpha(t) &=& \alpha_0 + \alpha_1 \,t \\
	   \delta(t) &=& \delta_0 + \delta_1 \,t\,.
		\label{eq:rotmod}
\end{eqnarray}
The rotation of Mercury around its axis is described by the longitude of the prime meridian $\varphi_0$, the rotation rate $\varphi_1$, and the physical longitudinal libration $\varphi_{\textrm{\small lib}}$
\begin{equation}
	   \varphi(t) = \varphi_0 + \varphi_1 t + \varphi_{\textrm{\small lib}}(t)\,.
\end{equation}
The rotation model is of great importance, as it is used to derive body-fixed coordinates of observations performed by spacecraft.
The matrix $\mathbf{R}$ which transforms coordinates from ICRF to body-fixed is composed from three rotations
\begin{equation}
  \mathbf{R}=\mathbf{R^{\textrm{\tiny T}}_z}(\varphi) \mathbf{R^{\textrm{\tiny T}}_x}(\pi/2-\delta)\mathbf{R^{\textrm{\tiny T}}_z}(\alpha+\pi/2)\,,
\end{equation}
where $\mathbf{R_{x,z}}$ denote counter-clockwise rotations (right hand rule) around the x- and z-axis, respectively. Note that the spin axis orientation can be computed by $\mathbf{e_s}= \mathbf{R_z}(\alpha+\pi/2)\mathbf{R_x}(\pi/2-\delta)\mathbf{e_Z}$.

\subsection{Resonant rotation}
Using the secular orbital elements in Tab. \ref{tab:elem} we derive the resonant spin rate of Mercury $\varphi_1^{(3/2)}$ for the case that the spin is in perfect 3:2 resonance to the motion of the planet on its orbit. Using the mean motion value $n_0 = M_1$ derived from the mean anomaly and the precession of the argument of pericenter $\omega_1$ we compute the mean resonant spin rate to
\begin{equation}
\varphi_1^{(3/2)} = \frac{3}{2} n_0 + \omega_1=(6.138506839\pm 0.000000028)
\,{}^\circ\!/\textrm{day}\,.
\label{eq:resrate}
\end{equation}
The current value of Mercury's rotation found in the literature is 6.1385025${}^\circ\!/\textrm{day}$ \citep{archinal11}. The difference between these rates corresponds to a longitudinal displacement of $5.7\,\textrm{arc sec}$ per year (approx. $67\,\textrm{m}$ per year at the equator of Mercury), which should be noticeable during e.g. the MESSENGER mapping mission (where typical image resolution vary from few kilometers to few meters).

We want to stress that the resonant spin rate $\varphi_1^{(3/2)}$ in the rotation model is composed of the planet's rotation around its spin axis and the precession of Mercury's orbit. Thereby, we have to consider the precession of the argument of pericenter $\omega_1=(5.164\pm 0.011)\times 10^{-6}\,{}^\circ\!/\textrm{day}$, and not the precession of the longitude of pericenter $\varpi_1=\Omega_1 + \omega_1$ since the precession of the ascending node $\Omega_1$ is already incorporated in the precession of the rotational axis. Note that the spin rate $\varphi_1^{(3/2)}$ is defined with respect to a precessing frame and is not strictly "sidereal" since the rotation axis changes slowly its orientation. Further, Mercury has a small but non-zero obliquity of $i_c=2.04\,\textrm{arc min}$ \citep{margot12}. However, the correction arising from the obliquity is in the order of $i_c\Omega_1^2/I_1$ and can be neglected when comparing with the error of $\omega_1$ (see appendix \ref{app:ic}).

If one of the sub-solar points at perihelion is used for the definition of the prime meridian $\varphi_0$, the orientation of Mercury's long axis at J2000.0 with resonant rotation would be
\begin{equation}
  \varphi^{(3/2)}_0 = \frac{3}{2}M_0 + \omega_0 = (329.7564\pm 0.0051){}^{\circ}\,.
\end{equation}
We find excellent agreement of this value with the findings of \cite{margot09}, who stated a value of $329.75{}^{\circ}$.
Note that the actual prime meridian of Mercury $\varphi_0$ is defined with respect to the crater Hun Kal located at $20{}^{\circ}\,\textrm{W}$ \citep{archinal11}.

\subsection{Physical libration in longitude}
The annual libration of Mercury is closely tied to the revolution of Mercury around the Sun. Of particular importance is the mean anomaly $M$ of Mercury since it defines the period and the phase of the libration. 
The libration is modeled as follows \citep{goldreich66}
\begin{equation}
\varphi_{\textrm{\small lib}}(t) = \sum_k g_{88/k}\sin\left(k\, n_0 (t+t_0)\right)\,.
	\label{eq:wlib}
\end{equation}
The amplitudes $g_{88/k}$ follow a recursive relationship
\begin{equation}
  g_{88/(k+1)}=g_{88/k}\frac{G_{2\,0\,1}(k+1,e_0)}{G_{2\,0\,1}(k,e_0)}\,,
\end{equation}
where $G_{2\,0\,1}(k,e_0)$ are given by Kaula's eccentricity functions \citep{kaula00}
\begin{equation}
  G_{2\,0\,1}(k,e)=\frac{G_{2\,0\,1-k}(e)-G_{2\,0\,1+k}(e)}{k^2}\,.
\end{equation}
Using $e_0=0.2056317\pm 0.0000071$ from Tab. \ref{tab:elem} we calculate the first five terms to 
\begin{eqnarray}
  G_{2\,0\,1}(1,e_0) = & 0.569650 &\pm 0.000027 \\
  G_{2\,0\,1}(2,e_0) = &(-60.0733 &\pm 0.0042)\times 10^{-3} \\
  G_{2\,0\,1}(3,e_0) = &(-5920.32 &\pm 0.77)\times 10^{-6} \\
	G_{2\,0\,1}(4,e_0) = &(-1200.10 &\pm 0.20)\times 10^{-6} \\
  G_{2\,0\,1}(5,e_0) = &(-267.691 &\pm 0.053)\times 10^{-6} \,.
\end{eqnarray}

The main period of the annual libration is the mean orbital period $T_{\textrm{\small orbit}}$ (Eq. \ref{eq:Torbit}) with the phase given by $M_0=n_0 t_0$. In addition, long-period variations of the orbital elements can lead to forced librational motion of Mercury with periods other than the orbital period \citep{peale07,yseboodt10}, but these are not considered in this work.

The measurement of the libration amplitude provides important constraints on the interior structure of Mercury. The amplitude of the annual libration $g_{88}$ is related by \citep{peale81}
\begin{equation}
  g_{88}=\frac{3}{2}\frac{B-A}{C_{m}}G_{2\,0\,1}(1,e_0)
\end{equation}
to the ratio of moments of inertia ${(B-A)}/{C_m}$, where $A\leq B<C$ are the principal axes of inertia of the planet and $C_m$ is the polar moment of inertia of the mantle and crust. Assuming the libration amplitude could be measured with a negligible error, the uncertainty in $(B-A)/C_m$ would be only at $6\times 10^{-7}$, due to the uncertainty in the eccentricity of Mercury's orbit. Here we used a libration amplitude of $g_{88}=38.5\,\textrm{arc sec}$ \citep{margot12}.

\subsection{Cassini state}
\label{sec:cas}
Mercury is assumed to occupy a Cassini state 1 \citep{peale69}, implying  that the spin vector of Mercury $\mathbf{e_s}$ lies in the plane defined by the Laplace plane normal $\mathbf{e_w}$ and the orbit normal $\mathbf{e_o}$ with the latter being enclosed by the others. The spin axis is consequently in a 1:1 resonance to the precession of the orbit normal, i.e. $\alpha_1 \approx \alpha_1^{\textrm{\small orbit}}$ and $\delta_1 \approx \delta_1^{\textrm{\small orbit}}$. Note that the spin axis precesses with slightly higher rates as described in appendix \ref{app:ic}.
The Cassini plane $\mathbf{e_c}$, which contains all Cassini states can be expressed as a linear combination of the orbit and Laplace plane normal
\begin{equation}
\mathbf{e_c} = r\,\mathbf{e_o} + s\,\mathbf{w}\,.
\end{equation}
We can constrain $r$ and $s$ by $|\mathbf{e_c}|=1$ and $\mathbf{e_c}\cdot \mathbf{e_o} = \cos i_c$, where $i_c$ is the obliquity. By using Eq. \ref{eq:w} for $\mathbf{w}$ and $\left|\mathbf{e_o} \times \mathbf{\dot{e}_o}\right|=\left|\mathbf{\dot{e}_o}\right|$ this results in 
\begin{equation}
\mathbf{e_c} = \cos i_c \mathbf{e_o} + \sin i_c
 \frac{\mathbf{e_o} \times \mathbf{\dot{e}_o}}{\left|\mathbf{\dot{e}_o}\right|}\,,
	\label{eq:ec}
\end{equation}
with 
\begin{eqnarray}
	\mathbf{e_o}\times\mathbf{\dot{e}_o} &=& \left(\dot{I}\cos\Omega-\dot{\Omega}\sin I\cos I\sin\Omega\right)\mathbf{e_X} \nonumber\\
	& & + \left(\dot{I}\sin\Omega+\dot{\Omega}\sin I\cos I\cos\Omega\right)\mathbf{e_Y} + \dot{\Omega}(\sin I)^2\mathbf{e_Z} \\
\left|\mathbf{\dot{e}_o}\right| &=& \mu\sin\iota = \sqrt{\dot{I}^2+(\dot{\Omega}\sin I)^2}\,.
	\label{eq:ecfull}
\end{eqnarray}

From Eq. \ref{eq:ec} it can be verified that the plane defining the Cassini state is independent from the exact form of the precession of the orbit around the Laplace plane. Especially, it is not dependent on $w_z$ as recognized by \citet{peale06} and \citet{yseboodt06}. The Cassini plane is sufficiently defined by the orientation of the orbit normal and its temporal change. In fact, the Cassini plane normal is the vector $\mathbf{\dot{e}_o}$ given by
\begin{eqnarray}
\mathbf{\dot{e}_o} &=& \left(\dot{\Omega}\cos\Omega\,\sin I + \dot{I}\sin\Omega\,\cos I \right)\,\mathbf{e_X} \nonumber\\
 &+& \left( \dot{\Omega}\sin\Omega\,\sin I - \dot{I}\cos\Omega\,\cos I \right)\,\mathbf{e_Y} \\
 &-& \dot{I}\sin I \,\mathbf{e_Z}\,.\nonumber
\end{eqnarray}

With the uncertainty of the orbital elements we can estimate the "thickness" of the Cassini plane, which results from uncertainties in the knowledge of the secular variation of Mercury's ephemeris. At $i_c=2.04\,\textrm{arc min}$ \citep{margot12} we find a $1\sigma$ thickness of $0.18\,\textrm{arc sec}$. This allows to interpret possible offsets of Mercury's spin orientation from the exact Cassini state \citep{margot12,peale14}.

Using the obliquity $i_c$ the polar moment of inertia $C/{m R^2}$ (scaled with the mass $m$ and radius $R$ of Mercury) can be calculated by \citep{peale81}
\begin{equation}
  \frac{C}{m R^2} = \frac{n_0\sin i_c((J_2 (1-e^2)^{-3/2}\cos i_c + C_{22}G_{201}(e)(1+\cos i_c))}{\mu \sin\iota\cos i_c-\mu\cos\iota\sin i_c}\,,
	\label{eq:cmr}
\end{equation}
where $J_2=5.03216\times 10^{-5}$ and $C_{22}=0.80389\times10^{-5}$ \citep{mazarico14} are the second degree harmonic coefficients of Mercury's gravity field. Assuming perfect knowledge in the obliquity and the gravitational coefficients the error on $C/{mR^2}$ is only at $6.1\times 10^{-5}$ due to the uncertainty of the orbital elements. Note that our analysis does not include model uncertainties of Eq. \ref{eq:cmr}, e.g., simplifying assumptions which were made in the derivation of the equation. We only infer the uncertainty of $C/{mR^2}$ due to orbital elements if Eq. \ref{eq:cmr} holds exactly and all other quantities are perfectly known. A more sophisticated analysis including higher order gravity field and tides can be found in \cite{noyelles13}.

\section{Discussion and conclusion}
In this work we extract orbital elements for Mercury from ephemeris data and predict a mean rotational model for Mercury in the view of a perfect resonance to its orbit. In this case the rotation is, besides the obliquity and the libration amplitude, completely determined by the mean orbital elements and their rates.
On the basis of the uncertainties in the mean orbital elements, errors of the theoretical perfectly resonant rotation model can be estimated. Note that ephemeris uncertainties are estimated from the periodic variation of the orbital elements and do not reflect any accuracy or "error" of the ephemeris. They can be rather understood as model uncertainties, since the secular part of Mercury's orbital elements does not capture the full variation of the orbit. In this work we introduced a consistent approach which allows us to estimate the uncertainties of the rotational parameters resulting from the simplified secular orbital elements. The findings are of great importance for interpretation of the current and future observations of Mercury's rotation by MESSENGER and BepiColombo spacecraft.


\section*{Appendix 1}
\label{app:ref}
The Keplerian orbital elements derived from the INPOP13c ephemeris \citep{fienga14} are given in Tab. \ref{tab:inpop}. We find very little difference of the values when comparing to the DE432 ephemeris (see Tab. \ref{tab:elem}). The deviation for the trend $a_1$ of the semi-major axis is about 1.7 meter per century.
\begin{table}[htb]
\begin{tabular}{c r r r }
\hline\noalign{\smallskip}
         & $x_0$ & $x_1$ & $x_2$ \\
				\noalign{\smallskip}\hline\noalign{\smallskip}
	  $a/(10^6\,\textrm{km})$ & $57.90909$ &     $-44\times 10^{-12}$ & $420\times 10^{-12}$ \\
		$e$  &      $0.2056317$ &    $20.4\times 10^{-6}$ &      $-0.027\times 10^{-6}$ \\
		$I/{}^{\circ}$  & $28.552197$ &     $0.0048473$ &    $-9.8\times 10^{-6}$ \\
		$\Omega/{}^{\circ}$ & $10.987969$ &     $-0.032808$ &   $-11.9\times 10^{-6}$ \\
		$\omega/{}^{\circ}$ & $67.5642$ &    $0.18862$ &   $  -4\times 10^{-6}$ \\
		$M/{}^{\circ}$  & $174.7948$ &     $149472.51578$ &   $   7\times 10^{-6}$ \\\hline
\end{tabular}
	\caption{The same as Tab. \ref{tab:elem} but derived from the INPOP13c ephemeris and a time span of 2000 years (09.06.973 AD - 23.06.2973).}
	\label{tab:inpop}
\end{table}

In Tab. \ref{tab:refs} we give values for reference frame dependent orbital elements with respect to the ecliptic (ECLIP, inclination $23.439291^{\circ}$), Mercury orbital plane (OP), and Mercury Laplace plane (LP) at J2000.0. The rotation matrices for the transformation to the these reference frames from the ICRF are given by
\begin{equation}
\mathbf{R}_{\textrm{\tiny ECLIP}}=\left(\begin{array}{r r r}
1  &  0  &  0 \\
0  &  0.91748206 &  0.39777716 \\
0  & -0.39777716 &  0.91748206 
\end{array}\right)\,,
\end{equation}
\begin{equation}
\mathbf{R}_{\textrm{\tiny OP}}=\left(\begin{array}{r r r}
 0.98166722  &   0.19060290  &  0 \\
-0.16742216  &   0.86227887  &  0.47795918\\
 0.09110040  &  -0.46919686  &  0.87838205
\end{array}\right)\,,
\end{equation}
\begin{equation}
\mathbf{R}_{\textrm{\tiny LP}}=\left(\begin{array}{r r r}
 0.88845611 &  0.43672271  &  0.14113473 \\
-0.45838720 &  0.82896828  &  0.32045711 \\
 0.02295468 & -0.34940643  &  0.93669004 
\end{array}\right) \,.
\end{equation}

The precession of the pericenter of Mercury is given by $\varpi_1^{\textrm{\tiny OP}}=\Omega_1^{\textrm{\tiny OP}}+\omega_1^{\textrm{\tiny OP}}=575.3\,\textrm{arc sec/cy}$. The inclination of the orbital plane with respect to the Laplace plane $I_0^{\textrm{\tiny LP}}=\iota=8.58^{\circ}$ remains constant $I_1^{\textrm{\tiny LP}}=I_2^{\textrm{\tiny LP}}\approx0$. The precession of the orbit around the Laplace plane is $|\Omega_1^{\textrm{\tiny LP}}|=\mu=0.109981^{\circ}/\textrm{cy}$.
 \begin{table}[htb]
\begin{tabular}{c r r r }
\hline\noalign{\smallskip}
         & $x_0$ & $x_1$ & $x_2$ \\
				\noalign{\smallskip}\hline\noalign{\smallskip}
				& \multicolumn{3}{c}{\textit{Ecliptic}}\\
		$I/{}^{\circ}$      
		                    &  $7.004975$ &  $0.0059524$ & $  0.7\times 10^{-6}$ \\ 
 		$\Omega/{}^{\circ}$ 
		                    & $48.330908$ & $-0.125416$  & $-89.2\times 10^{-6}$ \\
		$\omega/{}^{\circ}$ 
		                    & $29.1252$   &  $0.28428$   & $ 80\times 10^{-6}$   \\
		$\varpi/{}^{\circ}$	& $77.4561$   &  $0.15886$   & $-13\times 10^{-6}$\\\hline
												& \multicolumn{3}{c}{\textit{Mercury orbital plane}}\\
		$I/{}^{\circ}$      &  $0.0$      &  $-0.016413$ & $ -3.9\times 10^{-6}$ \\ 
 		$\Omega/{}^{\circ}$ & $68.735669$ & $-0.054375$  & $337.0\times 10^{-6}$ \\
		$\omega/{}^{\circ}$ &$320.3895$   &  $0.21417$   & $-350\times 10^{-6}$ \\
		$\varpi/{}^{\circ}$	& $29.1252$   &  $0.15980$   & $-13\times 10^{-6}$\\\hline
		                    & \multicolumn{3}{c}{\textit{Mercury Laplace plane}}\\
		$I/{}^{\circ}$      &  $8.582338$ &  $1\times 10^{-18}$       & $  5\times 10^{-21}$ \\ 
 		$\Omega/{}^{\circ}$ & $0.0$       & $-0.109981$  & $-25.9\times 10^{-6}$ \\
		$\omega/{}^{\circ}$ & $50.3895$   &  $0.26855$   & $ 13\times 10^{-6}$\\
		$\varpi/{}^{\circ}$	& $50.3895$   &  $0.15857$   & $-13\times 10^{-6}$\\\hline
  \end{tabular}
	\caption{Orbital elements of Mercury as derived from the DE432 ephemeris at epoch J2000.0 with respect to the following reference frames: Ecliptic and Earth equinox of J2000; Mercury orbital plane of J2000.0 and ascending node with respect to the ecliptic; Mercury Laplace plane and ascending node with respect to the Mercury orbital plane of J2000.0.}
	\label{tab:refs}
\end{table}

\section*{Appendix 2}
\label{app:freq}
In Tab. \ref{tab:freq} we list the first ten periodic terms, which were identified in the osculating orbital elements time series. Some of the periods can be assigned to planetary perturbations, e.g., Venus: $(\lambda_V)$ 0.62 years; $(2\lambda_V)$ 0.31 years; $(2\lambda_M-5\lambda_V)$ 5.66 years; $(\lambda_M - 3\lambda_V)$ 1.38 years; $(\lambda_M - 2 \lambda_V)$ 1.11 years; $(2\lambda_M-4\lambda_V)$ 0.55 years; Earth: $(\lambda_M-4\lambda_E)$ 6.58 years; Jupiter: $(\lambda_{J})$ 11.86 years; $(2\lambda_{J})$ 5.93 years; $(3\lambda_{J})$ 3.95 years; Saturn: $(2\lambda_{S})$ 14.73 years, where $\lambda=M+\varpi = M + \Omega + \omega$ denotes the mean longitude of the planet, respectively.

\begin{table}
\begin{tabular}{r r r r r r r r}
\hline\noalign{\smallskip}
&\multicolumn{3}{l}{$a$ [km]} & &
\multicolumn{3}{l}{$e$ [$10^{-6}$]} \\
  $i$ & $A_i$    &   $T_i$  & $\phi_i$ & &   $A_i$  &   $T_i$   & $\phi_i$  \\\hline
1 & 109.56 & 1.11 & 145.25 & & 7.23 &  5.93 & 272.97 \\
2 & 56.37  & 0.20 & 140.65 & & 4.63 &  1.11 & 326.10 \\
3 & 54.55  & 5.66 & 356.60 & & 3.53 &  5.66 & 181.05 \\
4 & 35.31  & 0.29 & 213.41 & & 2.53 &  1.38 & 141.21 \\
5 & 31.01  & 0.12 & 76.038 & & 1.52 & 11.86 & 123.83 \\
6 & 29.41  & 1.38 & 325.51 & & 0.89 &  0.25 & 261.39 \\
7 & 21.92  & 0.13 & 211.50 & & 0.88 &  0.55 & 109.00 \\
8 & 21.19  & 0.25 &  81.45 & & 0.86 & 14.73 & 304.98 \\
9 & 20.27  & 0.55 & 285.27 & & 0.81 &  0.46 & 128.90 \\
10 & 19.46 & 0.40 & 70.013 & & 0.78 & 0.29 &  36.27 \\\hline\noalign{\smallskip}
& \multicolumn{3}{l}{$I$ [$10^{-3}$ arc sec]} & & 
  \multicolumn{3}{l}{$\Omega$ [$10^{-3}$ arc sec]} \\
$i$ & $A_i$    &   $T_i$  & $\phi_i$ & &   $A_i$  &   $T_i$   & $\phi_i$  \\\hline
1 & 167.3 &  5.93 &  15.01 & & 399.8 &  5.93 & 292.75 \\
2 &  52.5 &  5.66 &  71.86 & & 165.8 &  5.66 & 135.33\\
3 &  31.9 &  1.38 & 250.97 & & 145.2 & 11.86 & 145.14\\
4 &  22.1 & 11.86 & 267.19 & & 116.6 &  1.11 & 249.76\\
5 &  20.1 & 14.73 &  58.16 & & 107.3 &  1.38 & 189.18\\
6 &  17.5 &  6.58 & 343.74 & &  57.5 &  0.62 & 356.75\\
7 &  17.2 &  3.95 &  51.01 & &  52.3 &  0.40 & 155.89\\
8 &  11.8 &  0.31 & 326.19 & &  48.3 & 14.73 & 336.52 \\
9 &   9.9 &  0.24 & 112.99 & &  44.0 &  6.56 &  59.20 \\
10 &  9.6 &  0.12 &  99.47 & &  42.6 &  0.24 &  45.75 \\\hline\noalign{\smallskip}
& \multicolumn{3}{l}{$\omega$ [arc sec]} & &
 \multicolumn{3}{l}{$M$ [arc sec]} \\
$i$ & $A_i$    &   $T_i$  & $\phi_i$ & &   $A_i$  &   $T_i$   & $\phi_i$  \\\hline
1 & 7.36 &  5.93 & 180.64 & & 10.71 &  5.66 &  87.97\\
2 & 4.57 &  1.11 &  55.93 & &  8.04 &  1.11 & 235.40\\
3 & 3.49 &  5.66 & 272.57 & &  7.70 &  5.93 &   3.50\\
4 & 2.55 &  1.38 &  50.09 & &  1.92 &  1.38 & 230.24\\
5 & 1.62 & 11.86 &  17.21 & &  1.90 & 11.86 & 186.46\\
6 & 0.89 & 14.73 & 212.80 & &  1.31 & 6.57 & 334.16\\
7 & 0.84 &  0.25 & 351.58 & &  1.22 & 0.55 &  17.86\\
8 & 0.84 &  0.55 & 200.04 & &  1.10 & 0.46 &  38.25\\
9 & 0.78 &  0.46 & 219.04 & &  1.10 & 0.29 & 305.42\\
10 & 0.78 & 0.62 & 194.46 & &  1.09 & 0.25 & 171.25
\\\hline
\end{tabular}
	\caption{Ten leading terms of the decomposition of the time series of the osculating orbital elements in $\sum_i A_i\cos(\nu_i\,t +\phi_i)$ with $\nu_i = 2\pi/T_i$. The unit of the amplitude $A_i$ is given in the brackets beside each orbital elements symbol, respectively. The periods $T_i$ are given in years and the phases $\phi_i$ in degrees. The values are given for orbital elements in the ICRF and derived from the DE432 ephemeris.}
	\label{tab:freq}
\end{table}

\section*{Appendix 3}
\label{app:ic}
The obliquity of the spin axis $i_c$ introduces small changes in the precession and resonant rotation rates. To stay within the Cassini plane the spin axis has to precess slightly faster than the orbital plane normal. In order to compute the corrections we expand equation Eq. \ref{eq:ec} to first order in the obliquity $i_c$. The declination $\delta$ and right ascension $\alpha$ of the spin axis are then given by
\begin{equation}
  \delta(t) = \frac{\pi}{2}-I + \frac{\dot{\Omega}\sin I}{\sqrt{\dot{I}^2+(\dot{\Omega}\sin I)^2}}i_c=\delta_0 + \delta_1 t
\end{equation}
and
\begin{equation}
  \alpha(t) = \Omega -\frac{\pi}{2}+ \frac{\dot{I}/\sin I}{\sqrt{\dot{I}^2+(\dot{\Omega}\sin I)^2}}i_c=\alpha_0 + \alpha_1 t\,.
\end{equation}
By deriving the series in $t$ we obtain the precession rates at J2000.0
\begin{equation}
  \delta_1 = I_1\left(-1 + \frac{\Omega_1 I_1^2 \cos I_0 + 2(\Omega_2 I_1-I_2\Omega_1)}{\sqrt{(I_1^2+(\Omega_1\sin I_0)^2)^3}}i_c\right)
\end{equation}
and
\begin{equation}
  \alpha_1 = \Omega_1 - \frac{(I_1^2\cot I_0 + \Omega_1^2\sin 2I_0)I_1^2/\sin I_0}{\sqrt{(I_1^2+(\Omega_1\sin I_0)^2)^3}}i_c\,.
\end{equation}
For $i_c=2.04\,\textrm{arc min}$ \citep{margot12} this results in
\begin{equation}
  \delta_1 = -0.00486^{\circ}/{\textrm{cy}} \textrm{ and }\alpha_1 = -0.03291^{\circ}/{\textrm{cy}}\,.
\end{equation}
The rotation rate is also slightly modified due to the obliquity. For small $i_c$ we get
\begin{equation}
  \varphi(t) = \frac{3}{2}M+ \omega - \frac{\dot{I}\cot I}{\sqrt{\dot{I}^2+(\dot{\Omega}\sin I)^2}}i_c=\varphi_0^{(3/2)}  + \varphi_1^{(3/2)}  t\,.
\end{equation}
The resonant rotation rate is consequently
\begin{equation}
  \varphi_1^{(3/2)}= \frac{3}{2}n_0 + \omega_1 + \\
	\frac{(I_1 \Omega_1)^2(3+\cos 2 I_0)/2 + (\Omega_2 I_1-I_2\Omega_1)\Omega_1\sin 2I_0+I_1^4/(\sin I_0)^2}{\sqrt{(I_1^2+(\Omega_1\sin I_0)^2)^3}}i_c
\end{equation}
and with $i_c=2.04\,\textrm{arc min}$ this amounts to $6.138506841^{\circ}/\textrm{day}$. The introduced correction is not significant when compared to the error of the resonant rotation rate in Eq. \ref{eq:resrate}.

\section*{Acknowledgments}
The authors thank Jean-Luc Margot, Beno\^it Noyelles, Stanton J. Peale, and Marie Yseboodt for fruitful discussions, and we also thank William M. Folkner and Charles H. Acton for providing information on the DE432 ephemeris. The reviews by two anonymous reviewers significantly improved earlier versions of this manuscript. J.O. greatly acknowledges being hosted by MIIGAiK and supported by the Russian Science Foundation under project 14-22-00197. The final publication is available at \href{http://dx.doi.org/10.1007/s10569-015-9633-4}{http://link.springer.com}

\bibliographystyle{spbasic}      
\bibliography{ref_141208}   

\end{document}